\documentclass[11pt]{article}

\usepackage[final]{acl}

\usepackage{times}
\usepackage{latexsym}

\usepackage[T1]{fontenc}

\usepackage[utf8]{inputenc}

\usepackage{microtype}

\usepackage{inconsolata}

\usepackage{graphicx}

\usepackage{booktabs}
\usepackage{multirow}
\usepackage{makecell}
\usepackage[most]{tcolorbox}
\usepackage{verbatim}
\usepackage{xspace}
\usepackage{url}
\usepackage{caption}

\newcounter{subfig}
\renewcommand{\thesubfig}{\alph{subfig}}

\usepackage{amsthm}
\theoremstyle{definition}
\newtheorem{definition}{Definition}
\DeclareMathOperator*{\argmin}{arg\,min}

\newtcolorbox{promptbox}[2][]{
  colback=blue!5!white,
  colframe=blue!75!black,
  fonttitle=\bfseries,
  title=#2,
  breakable,
  #1
}

\newcommand{\methodfont}[1]{\textsc{#1}}
\newcommand{\fullcode}{\methodfont{FullCode}\xspace}
\newcommand{\minunidiff}{\methodfont{MinUniDiff}\xspace}
\newcommand{\unidiff}{\methodfont{UniDiff}\xspace}
\newcommand{\mincontentdiff}{\methodfont{MinContentDiff}\xspace}
\newcommand{\contentdiff}{\methodfont{ContentDiff}\xspace}
\newcommand{\blockdiff}{\methodfont{BlockDiff}\xspace}
\newcommand{\funcdiff}{\methodfont{FuncDiff}\xspace}
\newcommand{\adaedit}{\methodfont{AdaEdit}\xspace}

\title{To Diff or Not to Diff? Structure-Aware and Adaptive Output Formats for Efficient LLM-based Code Editing}

\author{
    Wei Cheng$^1$\thanks{\,\,Work was partially done when interned at Tongyi Lab.}, 
    Yongchang Cao$^2$,
    Chen Shen$^1$,
    Binhua Li$^2$,
    Jue Chen$^2$,
    Yongbin Li$^2$\thanks{\,\,Corresponding authors.},
    Wei Hu$^1$\footnotemark[2] \\
    $^1$ State Key Laboratory for Novel Software Technology, Nanjing University, China \\
    $^2$ Tongyi Lab, Alibaba Group, China \\
    \texttt{\{wchengcs,cshen\}.nju@gmail.com, whu@nju.edu.cn} \\
    \texttt{\{caoyongchang.cyc,binhua.lbh,chenjue.chen,shuide.lyb\}@alibaba-inc.com} 
}

\begin{document}
\maketitle

\begin{abstract}
Large Language Models (LLMs) are increasingly used for code editing, yet the prevalent full-code generation paradigm suffers from severe efficiency bottlenecks, posing challenges for interactive coding assistants that demand low latency and cost.  
Despite the predominant focus on scaling model capabilities, the edit format itself has been largely overlooked in model training.
In this paper, we begin with a systematic study of conventional diff formats and reveal that fragile offsets and fragmented hunks make generation highly unnatural for LLMs.
To address it, we introduce \blockdiff and \funcdiff, two structure-aware diff formats that represent changes as block-level rewrites of syntactically coherent units such as control structures and functions. 
Furthermore, we propose \adaedit, a general adaptive edit strategy that trains LLMs to dynamically choose the most token-efficient format between a given diff format and full code. 
Extensive experiments demonstrate that \adaedit paired with structure-aware diff formats consistently matches the accuracy of full-code generation, while reducing both latency and cost by over 30\% on long-code editing tasks.
\end{abstract}

\section{Introduction}

Large Language Models (LLMs) \cite{deepseekcoder,qwencoder} have become a cornerstone of modern software engineering \cite{Dakhel2023GitHub,Kazemitabaar2023Studying}, where code editing that involves modifying source code according to an edit intent is one of fundamental tasks \cite{Lientz1978Characteristics,Fan2025Exploring}.
In Integrated Development Environments (IDEs), this task powers frequent, interactive features such as next edit suggestion \cite{cursor,Chen2025An}.
While the research community has predominantly focused on scaling model capabilities to push the ceiling of edit accuracy, the strict efficiency constraints of real-world deployments are often sidelined. 
For real-time collaborative workflows, low latency and cost-saving are not merely optional enhancements, but mandatory prerequisites \cite{Barke2023Grounded,Liang2024Large}.

Consequently, the predominant \emph{full-code generation} paradigm, where models always rewrite entire code for even minor edits, suffers from severe efficiency bottlenecks \cite{Guo2024CodeEditorBench,Singhal2024NoFunEval,Aggarwal2025NextCoder}. 
Even if the generation is accurate, this brute-force approach incurs high latency and prohibitive API costs by outputting vast amounts of redundant tokens, while simultaneously increasing the risk of unintended modifications \cite{Pereira2025Beyond}.
Therefore, resolving this efficiency bottleneck without sacrificing edit accuracy represents a critical research challenge with immense industrial value.

\begin{figure*}[!t]
\centering
\begin{minipage}[t]{0.4\textwidth}
    \centering
    \refstepcounter{subfig}
    \includegraphics[width=\linewidth]{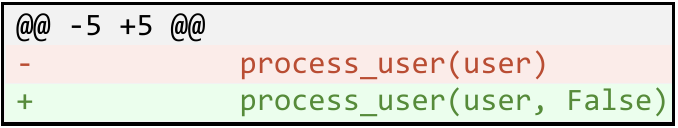}
    \caption*{(\thesubfig) An example of \minunidiff.}
    \label{fig:example-minunidiff}
    \vspace{0.5em}
    \setcounter{subfig}{2}
    \refstepcounter{subfig}
    \includegraphics[width=\linewidth]{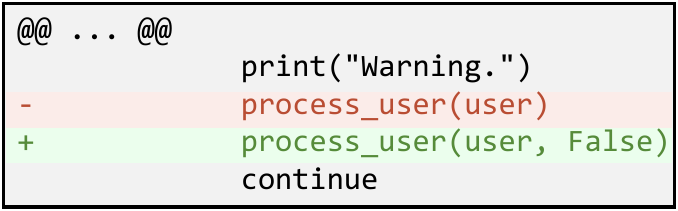}
    \caption*{(c) An example of \mincontentdiff.}
    \label{fig:example-mincontentdiff}
\end{minipage}
\hspace{1em}
\begin{minipage}[t]{0.4\textwidth}
    \centering
    \setcounter{subfig}{1}
    \refstepcounter{subfig}
    \includegraphics[width=\linewidth]{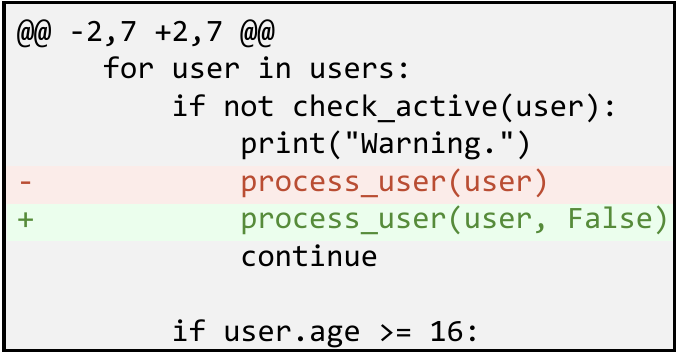}
    \caption*{(b) An example of \unidiff. The diff content of \contentdiff is identical, but the line number header is replaced with \texttt{@@ ... @@}.}
    \label{fig:example-unidiff}
\end{minipage}
\caption{Examples of conventional diff formats. Refer to Figure~\ref{fig:example} for the complete code editing task.}
\label{fig:example-diffs}
\end{figure*}

Despite this pressing need, the \emph{edit format} itself, i.e., the representation of code changes, has been largely overlooked as a pathway to efficiency.
Current models either output full code or exhibit lazy coding by using informal placeholders, which lack the precision required for automated patching.
Existing explorations \cite{aider2023edit} of alternative edit formats, such as variants of the unified diff \cite{unidiff} shown in Figure~\ref{fig:example-diffs}, are confined to prompting strategies for powerful instruction-following models.
Consequently, the fundamental question of \emph{how to design and adapt edit formats} in LLM-based code editing remains open.

In this paper, we first conduct a systematic study of conventional diff formats.
Our findings reveal that number-indexed diff formats are highly fragile due to precise numerical offsets, while content-addressed diff formats mainly suffer from fragmented hunks that break the syntactic integrity of code.
Generating these formats are fundamentally mismatched with the generative nature of LLMs, leading to substantially degraded edit accuracy.

To respond to our findings, we introduce \blockdiff and \funcdiff, two structure-aware diff formats that represent changes as block-level rewrites of syntactically coherent units, such as control structures and functions.
By aligning textual diffs to code blocks extracted from the Abstract Syntax Tree (AST), we enable LLMs to edit in logical modules rather than arbitrary line fragments, which restores the naturalness of the generation process.
However, any edit format is not universally optimal, where the overhead of diff formats can exceed that of full-code generation when changes are pervasive across the code.
We therefore propose \adaedit, a general \underline{ada}ptive \underline{edit} strategy that empowers LLMs to dynamically choose between a given diff format and full code. 
The model internalizes this switching logic during finetuning by trained on the most token-efficient 
format for each individual sample.

We evaluate different edit formats on diverse datasets \cite{ocedata,instructcoder,humanevalfix,canitedit,aider} and base models \cite{deepseekcoder,qwencoder}.
Our results demonstrate that \adaedit paired with structure-aware diff formats consistently matches full-code generation in edit accuracy, while reducing both latency and cost by over 30\% on long-code editing tasks. 

In summary, our main contributions are outlined as follows:
\begin{itemize}
    \item We conduct a systematic study of conventional diff formats, revealing their fragile offsets and fragmented hunks for unnatural generation.
    
    \item We introduce two structure-aware diff formats that represent changes as block-level rewrites of syntactically coherent units.
    
    \item We propose \adaedit, a general adaptive edit strategy that trains LLMs to choose the most token-efficient format between a given diff format and full code.
    
    \item We conduct extensive experiments to demonstrate that \adaedit paired with structure-aware diff formats guarantees edit accuracy while significantly reducing latency and cost over full-code generation.
    Our source code is available at \url{https://github.com/nju-websoft/AdaEdit}.
\end{itemize}

\begin{table*}
    \centering
    {\small
    \begin{tabular}{l|ccccc|c}
    \toprule
    
    Formats & EditEval & CanItEdit  & HumanEvalFix  & Aider-1  & Aider-2  & Average \\
    
    \midrule
    
    Base model          & 55.39 & 42.98  & 65.12  & 35.56  & 44.44  & 48.70 \\
    \fullcode           & 69.38 & 53.17  & 65.76  & 45.93  & 51.11  & 57.07 \\

    \midrule

    \minunidiff         & 21.44 & \ \ 7.60  & 12.41  & 13.33  & 15.56  & 14.07 \\
    \quad w/ numbers    & 41.49 & 23.93  & 42.80  & 20.74  & 26.67  & 31.13 \\
    
    \unidiff            & 48.07 & 17.33  & 39.60  & 28.89  & 31.85  & 33.15 \\
    \quad w/ numbers    & 51.65 & 30.67  & 41.55  & 28.89  & 35.56  & 37.66 \\

    \midrule

    \mincontentdiff     & 61.42 & 37.02  & 52.20  & 35.56  & 42.22  & 45.68 \\
    \contentdiff        & 67.91 & 47.19  & 65.91  & 43.70 & 47.41 & 54.43 \\

    \bottomrule
    \end{tabular}}

    \caption{Pass@1 comparison of conventional diff formats, trained on the Qwen2.5-Coder-7B base model.}
    \label{tab:tra-exp}
\end{table*}

\section{Related Work}

\subsection{Code Editing with LLMs}

Automated code editing initially focus on localized tasks like code completion and infilling \cite{Izadi2022CodeFill, Fried2023InCoder}, which leverage Fill-in-the-Middle (FIM) techniques to predict missing regions \cite{Bavarian2022Efficient}. 
More recently, specialized models have emerged to perform non-trivial modifications without explicit location hints \cite{Zhang2022CoditT5, Li2023CodeEditor}.
The advent of modern LLMs \cite{deepseekcoder,gpt4,qwencoder,gemini} has further unified these disparate tasks under a single generative framework, giving rise to what we term general-purpose code editing.
It is defined as the task of modifying source code according to an edit intent, which may be expressed as explicit natural language instructions \cite{Guo2024CodeEditorBench,Singhal2024NoFunEval,canitedit}, or implicit goals such as code repair and efficiency improvements \cite{Zhang2023selfedit,Joshi2023Repair,Wei2023Copiloting,Chen2024Teaching,Zhang2024Pair,Olausson2024Is}.
Across this spectrum, a common paradigm of full-code generation has been adopted, leading most existing work to focus on enhancing the models' intrinsic capabilities to improve edit accuracy \cite{humanevalfix,instructcoder,Aggarwal2025NextCoder,ocedata}.
However, the fundamental role of the output format in determining edit efficiency has been largely overlooked.

\subsection{Diff Formats and Their Applications}

Text-based diff formats identify changes between plain text blocks and serve as the foundation of modern version control \cite{Hunt1976algorithm, Myers1986An}.
However, the line-number indexing in these formats \cite{contextdiff,normaldiff,unidiff} is inherently fragile for LLM generation since probabilistic models struggle with precise numerical offsets \cite{codegendiff,humanevalfix}.
Alternatively, syntax-aware diff tools like GumTree represent modifications as symbolic operations on ASTs \cite{Fluri2007Change, falleri2014fine, Martinez2023Hyperparameter, Falleri2024Fine}.
While effective for code analysis, these formats require specialized domain-specific languages that are fundamentally misaligned with the sequential text generation paradigm of LLMs.
Furthermore, they typically ignore changes to non-semantic nodes, discarding modifications to spaces and even comments. 
This characteristic makes them entirely unsuitable for precise code generation and textual reconstruction.
Recent studies \cite{aider2023edit,Pereira2025Beyond} have explored content-addressed diff formats through sophisticated prompt engineering, relying on the powerful instruction-following capabilities of LLMs.
In contrast, our work provides the first systematic training-based investigation into edit formats.
Moreover, we operate under a completely different paradigm with existing diff tools, where we capture all textual modifications based on the standard unified diff and expand these exact diff hunks to encompass code structures.
This ensures absolute textual fidelity while still leveraging structural context, culminating in the proposal of novel structure-aware formats and an adaptive edit strategy.

\section{Challenges in Learning Diff Formats}

\subsection{Problem Formulation}

To systematically investigate edit formats, we formalize LLM-based code editing as follows.
Let $\mathcal{I}$ be the space of edit intents, $\mathcal{C}$ be the space of code snippets, and $\mathcal{E}$ be the space of edit representations.

\begin{definition}[Edit Format]
An edit format is a pair of complementary functions, $\text{Diff}: \mathcal{C} \times \mathcal{C} \rightarrow \mathcal{E}$ and $\text{Patch}: \mathcal{C} \times \mathcal{E} \rightarrow \mathcal{C}$. 
For any source code $C \in \mathcal{C}$ and target code $C' \in \mathcal{C}$, these functions must satisfy the reconstruction identity:
\begin{equation}
\label{eq:format}
    \text{Patch}\left(C, \text{Diff}(C, C')\right) = C',
\end{equation}
In this framework, full-code generation is a special case where $\text{Diff}(C, C') = C'$, while the $\text{Patch}$ function serves as an identity projection.
\end{definition}

\begin{definition}[Edit Format Learning]
The goal of learning an edit format is to train a generative model $\mathcal{M}_{\theta}$ using a dataset $\mathcal{S} = \{(I_j, C_j, C'_j)\}_{j=1}^n$. 
Each triplet consists of an intent $I_j \in \mathcal{I}$, a source code $C_j \in \mathcal{C}$, and a target code $C'_j \in \mathcal{C}$. 
The learning process first transforms $\mathcal{S}$ into a training set $\mathcal{D} = \{(I_j, C_j, E_j)\}_{j=1}^n$ by computing $E_j = \text{Diff}(C_j, C'_j)$. 
Then, the model parameters are optimized by minimizing the objective:
\begin{equation}
    \theta^* = \argmin_{\theta} \sum_{j=1}^{n} \mathcal{L}\left(\mathcal{M}_{\theta}(I_j, C_j), E_j\right),
\end{equation}
where $\mathcal{L}$ represents the token-level cross-entropy loss. 
During inference, $\mathcal{M}_{\theta^*}$ generates the specific edit representations for subsequent patching.
\end{definition}

\begin{figure*}
    \centering
    \includegraphics[width=\textwidth]{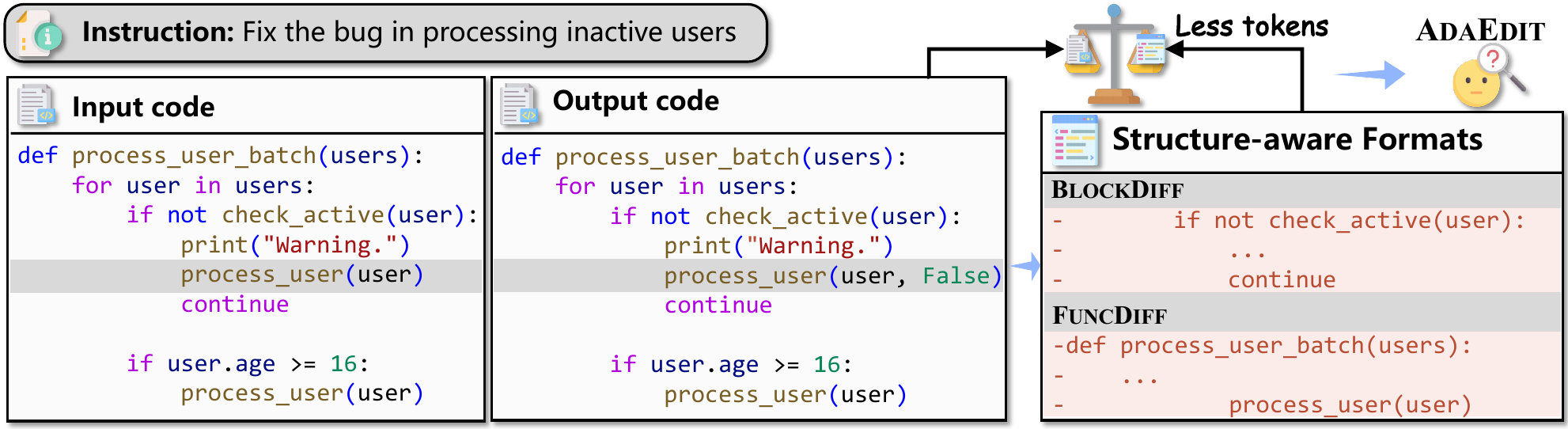}
    \caption{Overview of structure-aware diff formats and \adaedit. Due to space constraints, lengthy code structures and added lines (+) are omitted in the diff examples.}
    \label{fig:example}
\end{figure*}

\subsection{Number-Indexed Diff Formats}
\label{sec:number}

Following the experiment setup detailed in Section~\ref{sec:setup}, we first investigate the widely-used unified diff format \cite{unidiff} that relies on line numbers to locate changes. 
To improve robustness against malformed model outputs, we implement a custom patch tool that bypasses the standard consistency verification and operates solely on the specified line numbers.
It includes two configurations:
\begin{itemize}
    \item \textbf{\minunidiff} is the minimal unified diff without context lines, as shown in Figure~\ref{fig:example-diffs}\ref{fig:example-minunidiff}.
    
    \item \textbf{\unidiff} is the standard unified diff with three context lines, as shown in Figure~\ref{fig:example-diffs}\ref{fig:example-unidiff}.
\end{itemize}

For both formats, we also evaluate a ``w/ numbers'' variant, where the source code is augmented with line numbers to help LLM generation.

As depicted in Table~\ref{tab:tra-exp}, all number-indexed diff formats yield edit accuracies far below that of the full-code baselines \cite{codegendiff,humanevalfix}. 
This poor performance stems from the inherent fragility: LLMs struggle to generate precise line numbers and offsets, and this issue persists even when the input code is explicitly numbered.

\subsection{Content-Addressed Diff Formats}
\label{sec:content}

To overcome the fragility of precise numerical offsets, content-addressed diff formats identify each edit region by a unique anchor content.
They differ in the anchor content and follow a unified patching process detailed in Section~\ref{sec:blockdiff}.
In this section, we include two variants:
\begin{itemize}
    \item \textbf{\mincontentdiff} is distinguished by its minimal anchor content. 
    Each hunk's anchor is constructed by iteratively adding one surrounding context lines at a time, both above and below the edit position, until the resulting content is unique within the source code.
    An example is shown in Figure~\ref{fig:example-diffs}\ref{fig:example-mincontentdiff}.

    \item \textbf{\contentdiff} is based on \mincontentdiff but requires as least three context lines, as shown in Figure~\ref{fig:example-diffs}\ref{fig:example-unidiff}.
\end{itemize}

We also investigate different hunk styles in Appendix~\ref{appendix:style}, such as the unified diff-like and search/replace style \cite{aider2023edit}.
Considering accuracy and robustness, we adopt the hunk rewrite style for all content-addressed diff formats.

The results in Table~\ref{tab:tra-exp} reveal a crucial insight. 
While \mincontentdiff outperforms all number-indexed diff formats, its accuracy remains substantially lags the full-code baselines. 
This indicates that eliminating line numbers is a necessary but insufficient step towards an effective edit format. 
While \contentdiff demands higher computational cost, this investment enhances the robustness of the anchor content, thereby significantly improving edit accuracy.
Despite these gains, it still falls short of the finetuned full-code baseline.

We attribute the remaining performance gap to a form of fragmented hunks.
In these conventional diff formats, the hunks consist of arbitrary line fragments.
For an LLM extensively pre-trained on well-formed code, generating such disjointed snippets is fundamentally unnatural.

\section{Methodology}


The contrasting performance between full-code generation and conventional diff formats reveals a fundamental tradeoff between efficiency and naturalness. 
Full-code generation is inherently natural for LLMs since it aligns with their pre-training objectives, yet it suffers from extreme redundancy by re-generating unchanged code. 
In contrast, conventional diff formats aim to reduce token counts by targeting only modified segments, but their fragmented hunks are unnatural, which frequently compromises edit accuracy.
Therefore, the central challenge lies in designing an edit format that achieves better efficiency than full-code generation while preserving sufficient naturalness for accuracy.

\subsection{Structure-Aware Diff Formats}
\label{sec:blockdiff}

Unlike line-level changes in natural languages, code changes typically fall into syntactically cohesive units.
Therefore, we introduce \blockdiff and \funcdiff, two structure-aware diff formats that represent changes as block-level rewrites, as shown in Figure~\ref{fig:example}.
By operating at different block levels, we hypothesize that they can restore sufficient naturalness for LLM generation.

\paragraph{Block Tree Construction}
We use \texttt{tree-sitter} \cite{treesitter} to construct an AST from the source code. 
By identifying fine-grained AST nodes, we build a block tree including control structures (e.g., branches, loops, and contextual blocks) and functions.
This hierarchy also incorporates coarse-grained nodes including classes and a \texttt{root} node representing the entire code.
To maintain compatibility with unstructured segments and malformed code snippets, any contiguous lines outside fine-grained nodes, such as \texttt{import} statements and regions with syntax errors, are encapsulated into special fine-grained nodes. 
This strategy prevents local modifications from being improperly promoted to a coarse-grained parent.
Based on this hierarchical structure, we define two distinct edit granularities. 
\blockdiff permits edits at any fine-grained node level to maximize token efficiency, while \funcdiff ignores control structures to favor broader structural stability.

\paragraph{Diff Generation}
Given a source code and a target code, we first compute their \minunidiff to cover all text diffs. 
Each diff hunk is then mapped to the block tree. 
A pure-insertion hunk is mapped to the smallest node that contains the insertion position. 
For a hunk that includes deletions, which may span multiple blocks, the target is the smallest set of contiguous block nodes that collectively contain all the deleted lines.

Although code blocks represent distinct units, their textual content is not always unique within the source code. 
To ensure unambiguous patching, we progressively expand the anchor content until it reaches contextual uniqueness.
This process first incorporates immediately adjacent sibling nodes and then expands to the parent node if ambiguity persists.
For each diff hunk, the expansion iterates until the selected anchor content is unique. 
In extreme cases where local uniqueness cannot be established, the expansion continues to the root node, thereby treating the entire code as the anchor.

Following the mapping and anchor expansion phases, individual hunks may contain overlapping lines. 
To eliminate redundant anchor content, we first merge any hunks that exhibit line-level overlaps. 
Furthermore, multiple distinct hunks may fall within the scope of a single fine-grained node, e.g., two branch changes in a loop. 
To enhance the naturalness of the generative process, we consolidate these hunks into their shared parent node. 
This transformation presents the modifications as a single logical edit instead of a series of fragmented changes. 
This structural merging proceeds iteratively in a bottom-up manner to ensure coherence across the entire block tree.

\paragraph{Patching}
The patching process is purely textual search and replacement operations without AST parsing.
Given a source code and a content-addressed diff, each hunk is interpreted as locating the code region by its anchor content, then performing the specified deletion and insertion operations to transform the code. 
A patch failure occurs if the anchor cannot be uniquely located, resulting from either zero matches or multiple ambiguous matches.
To mitigate occasional non-determinism in LLM outputs, the process incorporates a tolerance mechanism. 
It attempts to find a plausible match by progressively relaxing constraints on whitespace and blank lines when an exact textual match is unavailable \cite{aider2023edit}.

\begin{table*}
    \centering
    {\small
    \begin{tabular}{c|l|ccccc|c}
    \toprule
    
    Base models & Formats & EditEval & CanItEdit & HumanEvalFix & Aider-1 & Aider-2 & Average \\
    
    \midrule

    \multirow{8}{*}{\makecell{DeepSeek-\\Coder-6.7B}} 
    & Base model              & 40.03 & 29.62  & 42.65  & 27.41  & 37.04  & 35.35 \\
    & \fullcode         & 64.38 & \textbf{44.88}  & 56.95  & \underline{45.19}  & 49.63  & \underline{52.21} \\
    \cmidrule{2-8}
    & \contentdiff          & 60.95 & 36.74  & 55.79  & 42.22  & 48.89  & 48.92 \\
    & \quad w/ \adaedit     & 63.84 & 36.19  & 59.39  & 42.96  & 49.63  & 50.40 \\
    
    & \blockdiff        & \textbf{64.87} & 34.74  & 58.02  & 44.44  & 51.11  & 50.64 \\
    & \quad w/ \adaedit   & 64.69 & \underline{38.98}  & \textbf{60.09}  & \underline{45.19}  & \underline{51.85}  & 52.16 \\
    
    & \funcdiff         & 64.25 & 35.86  & \underline{59.79}  & 44.44  & 49.63  & 50.79 \\
    & \quad w/ \adaedit   & \underline{64.79} & 38.95 & 59.76 & \textbf{46.67} & \textbf{52.59} & \textbf{52.55} \\

    \midrule
    
    \multirow{8}{*}{\makecell{Qwen2.5-\\Coder-7B}} 
    & Base model              & 55.39 & 42.98  & 65.12  & 35.56  & 44.44  & 48.70 \\
    & \fullcode         & 69.38 & \textbf{53.17}  & 65.76  & \underline{45.93}  & 51.11  & 57.07 \\
    \cmidrule{2-8}
    & \contentdiff        & 67.91 & 47.19  & 65.91  & 43.70  & 47.41  & 54.43 \\
    & \quad w/ \adaedit   & 68.02 & 48.60  & 65.88  & 44.44  & 49.63  & 55.31 \\
    
    & \blockdiff        & \underline{69.95} & 48.07  & 64.85  & 45.19  & \underline{51.85}  & 55.98 \\
    & \quad w/ \adaedit   & 69.30 & 51.36  & 67.38  & \textbf{47.41}  & \textbf{52.59}  & \underline{57.61} \\
    
    & \funcdiff         & \textbf{70.88} & 50.31  & \underline{67.62}  & \underline{45.93}  & \underline{51.85}  & 57.32 \\
    & \quad w/ \adaedit   & 69.23 & \underline{52.67} & \textbf{67.84} & \textbf{47.41} & \textbf{52.59} & \textbf{57.95} \\

    \midrule
    
    \multirow{8}{*}{\makecell{Qwen2.5-\\Coder-14B}} 
    & Base model              & 58.25 & 46.67  & 71.52  & 41.48  & 51.11  & 53.81 \\
    & \fullcode         & 69.59 & \textbf{60.95}  & 70.40  & \underline{54.07}  & \textbf{64.44}  & 63.89 \\
    \cmidrule{2-8}
    & \contentdiff          & \underline{72.40} & 57.69  & 66.65  & 53.33  & 60.74  & 62.16 \\
    & \quad w/ \adaedit     & 70.21 & 55.07  & 69.18  & \underline{54.07}  & 62.22  & 62.15 \\
    
    & \blockdiff        & \underline{72.40} & 60.40  & 69.97  & \underline{54.07}  & \underline{63.70}  & 64.11 \\
    & \quad w/ \adaedit   & 71.47 & \underline{60.48}  & 69.88  & \underline{54.07}  & \underline{63.70}  & 63.92 \\
    
    & \funcdiff         & \textbf{73.48} & 58.64  & \textbf{73.05}  & \textbf{54.81}  & \textbf{64.44}  & \textbf{64.89} \\
    & \quad w/ \adaedit   & 71.52 & 60.21  & \underline{72.41}  & \textbf{54.81}  & \textbf{64.44}  & \underline{64.68} \\

    \bottomrule
    \end{tabular}}

    \caption{Main comparison of pass@1.
    The best results are marked in \textbf{bold}, and the second best are \underline{underlined}.}
    \label{tab:main-exp}
    
\end{table*}

\subsection{\adaedit: An Adaptive Strategy}

The efficiency of diff formats is intrinsically tied to the magnitude of changes within the source code.
Since these representations require both anchor and modified content, the token count can escalate when edits are extensive or dispersed across multiple regions. 
Under such conditions, the cumulative overhead of diff hunks may eventually exceed the cost of full-code regeneration.
This creates a critical threshold where a diff format is no longer the most efficient choice for code editing, effectively negating the efficiency benefits typically associated with localized editing.

To ensure optimal efficiency across all scenarios, we propose \adaedit, a general adaptive edit strategy that trains LLMs to dynamically choose the most token-efficient format between a given diff format and full code. 
This switching logic is internalized during training through a data-driven proxy.
For each pair of source code $C_j$ and target code $C'_j$ in the training set, we compare the lengths of the diff representation and the full target code. 
We then re-define the ground truth $E_j$ as the shorter format for that instance:
\begin{equation}
    E_j = \argmin_{S \in \{C'_j, \text{Diff}(C_j, C'_j)\}} |S|,
\end{equation}
where $|\cdot|$ denotes the token count operator defined by the specific tokenizer of LLMs. 
By finetuning on this optimized dataset, the model implicitly learns to predict the more efficient format at inference time based on the provided editing task.

\section{Experiments and Results}
\label{sec:setup}

\subsection{Datasets}

Given the availability of high-quality code editing resources, we primarily focus on Python.
Our main training dataset is \textbf{OCEData} \cite{ocedata}.
We evaluate on four diverse Python benchmarks. 
These include two instruction-guided benchmarks that aligns with the training task: \textbf{EditEval} \cite{instructcoder} and \textbf{CanItEdit} \cite{canitedit}. 
We also incorporate two bug-fixing benchmarks that involve implicit edit intents: \textbf{HumanEvalFix} \cite{humanevalfix}, a test-driven repair task, and \textbf{Aider} \cite{aider}, which simulates realistic two-stage coding exercises involving initial generation and subsequent test-driven repair. 

To demonstrate the generalizability, we also evaluate on another Python training dataset \textbf{InstructCoder} \cite{instructcoder}. 
For other programming languages, we train on the JavaScript subset of \textbf{CommitPackFT} \cite{humanevalfix} and evaluate on HumanEvalFix-JavaScript.
More details are presented in Appendix~\ref{appendix:datasets} and Table~\ref{tab:dataset-stats}.

\subsection{Implementation Details}

We compare diff formats against two full-code baselines: the original non-finetuned \textbf{Base} model, and \textbf{\fullcode} that serves as the upper bound for naturalness and accuracy.
The representative conventional format \contentdiff is also included for comparison with our structure-aware formats.

Our experiments primarily utilize the Qwen2.5-Coder series \cite{qwencoder}, specifically 7B and 14B, which represent the state-of-the-art in open-source code LLMs. 
We also evaluate on DeepSeek-Coder-6.7B \cite{deepseekcoder}, an earlier model with different architectural variants.
All models undergo full-parameter Supervised Fine-Tuning (SFT) from their base versions to ensure that performance variations stem solely from the choice of edit format.
We apply a unified prompt template with format-specific prefixes across all evaluations.
More details are shown in Appendix~\ref{appendix:impl}.

\subsection{Evaluation Metrics}

We assess each edit format from two primary perspectives and report the average of each metric.
See Appendix~\ref{appendix:metric} for more details:
\begin{itemize}
    \item \textbf{Effectiveness.} We evaluate edit accuracy by pass@1, and usability by exact patch-apply success rate and stricter linter check.

    \item \textbf{Efficiency.} We evaluate edit latency by tokens of the first renderable generation, and cost by tokens of the complete generation.
\end{itemize}

For \adaedit, we further report the percentage of samples where the model correctly chooses the most token-efficient edit format.

\begin{figure}
    \centering
    \includegraphics[width=\linewidth]{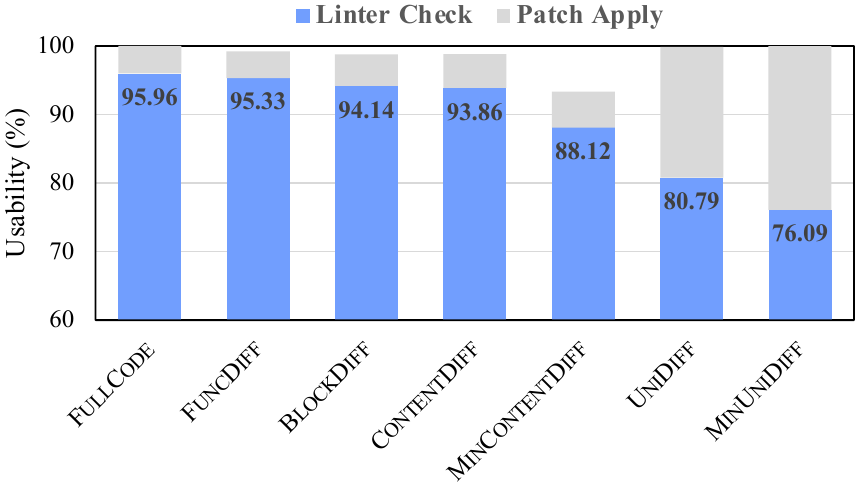}
    \caption{Edit usability comparison, trained on Qwen2.5-Coder-7B. 
    The source code for \minunidiff and \unidiff is augmented with line numbers.}
    \label{fig:usability}
\end{figure}

\subsection{Effectiveness: Edit Accuracy \& Usability}
\label{sec:effect}

The main results in Table~\ref{tab:main-exp} demonstrate that \blockdiff and \funcdiff consistently outperform the line-level \contentdiff across all base models.
By preserving syntactic coherence of diff hunks, structure-aware formats significantly restore the naturalness for LLM generation.
We observe that \funcdiff consistently achieves superior accuracy among diff formats. 
With the more capable Qwen2.5-Coder models, the average accuracy of \funcdiff even exceeds the \fullcode baseline. 
A similar phenomenon occurs with \blockdiff when trained on Qwen2.5-Coder-14B, which suggests that the advantages of structural representations require more powerful models to fully realize.

The integration of \adaedit further enhances accuracy by enabling models to dynamically choose between a specific diff format and full-code generation.
Although \adaedit is specifically optimized for token efficiency, the reduction in test-time inference costs does not generally compromise edit accuracy. 
This adaptivity is especially effective for complex tasks such as CanItEdit and Aider, demonstrating that a purely diff-based approach is not always the optimal solution. 
For Qwen2.5-Coder-14B, its superior capacity allows for generating effective diff representations even in complex tasks while fully leveraging test-time scaling on function-level benchmarks like EditEval \cite{Hoffmann2022Training}. 
In such instances, \adaedit maintains accuracy while significantly reducing inference costs, as evidenced in Section~\ref{sec:efficiency}. 
Furthermore, structure-aware diff formats still consistently outperform \contentdiff when integrated with \adaedit.

\begin{figure}
    \centering
    \includegraphics[width=\linewidth]{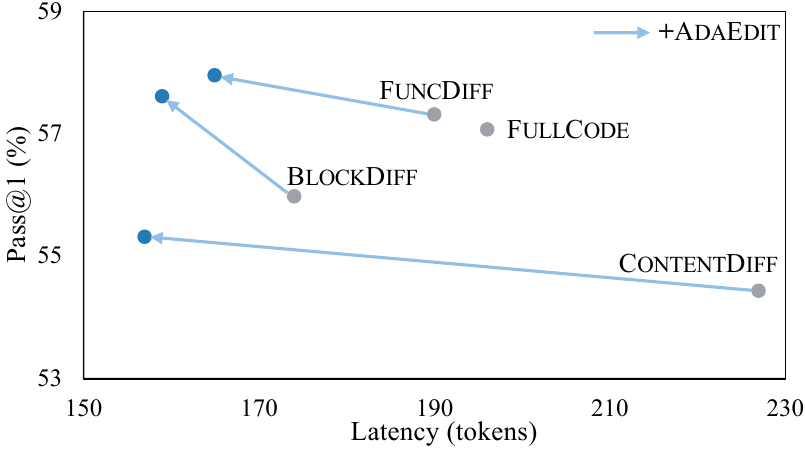}
    \caption{Latency-accuracy landscape of different edit formats and \adaedit, trained on Qwen2.5-Coder-7B.}
    \label{fig:cost-effect}
\end{figure}

\begin{figure}
    \centering
    \includegraphics[width=\linewidth]{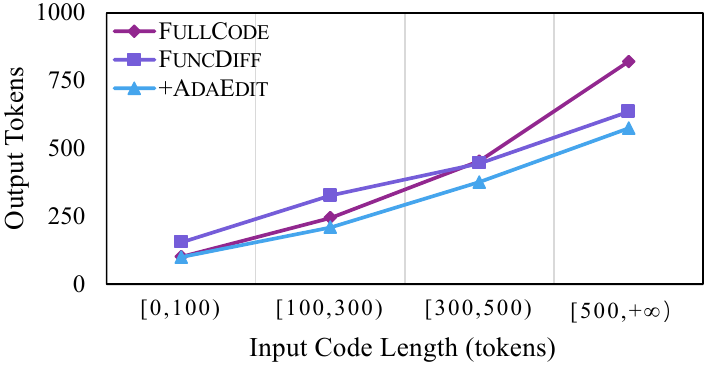}
    \caption{Edit cost comparison across code scales, trained on Qwen2.5-Coder-7B.}
    \label{fig:cost}
\end{figure}

Except for functional correctness, we assess the basic usability of generated edits, as illustrated in Figure~\ref{fig:usability}. 
Only \contentdiff frequently encounters patch application failures, since the model is more likely to generate minimal anchor content that is not unique within the source code.
In contrast, while number-indexed formats maintain nearly perfect patch success rates due to the absence of strict consistency verification, subsequent linter checks reveal significant damage to the resulting code. 
When evaluated under stricter linter checks, the observed usability trends closely mirror edit accuracy, where structure-aware formats consistently achieve superior usability among diff formats.

\subsection{Efficiency: Edit Latency \& Cost}
\label{sec:efficiency}

\begin{table}
    \centering
    {\small
    \begin{tabular}{l|cc}
    \toprule
    
    Formats & Pass@1 (\%) & Cost (tokens) \\
    \midrule

    \fullcode           & 39.75 & 648.30 \\
    \midrule
    
    \contentdiff        & 33.75 & 612.85 \\ 
    \quad w/ \adaedit   & 33.00 & \textbf{432.73} \\ 
    \blockdiff          & 38.69 & 570.26 \\ 
    \quad w/ \adaedit   & 37.94 & \underline{466.04} \\  
    \funcdiff           & \textbf{40.75} & 546.77 \\ 
    \quad w/ \adaedit   & \underline{40.69} & 481.63 \\ 

    \bottomrule
    \end{tabular}}
    \caption{Performance comparison on the CanItEdit subset containing 80 samples whose input code length is over 300 tokens, trained on Qwen2.5-Coder-7B.}
    \label{tab:canitedit}
\end{table}

\begin{figure}
    \centering
    \includegraphics[width=\linewidth]{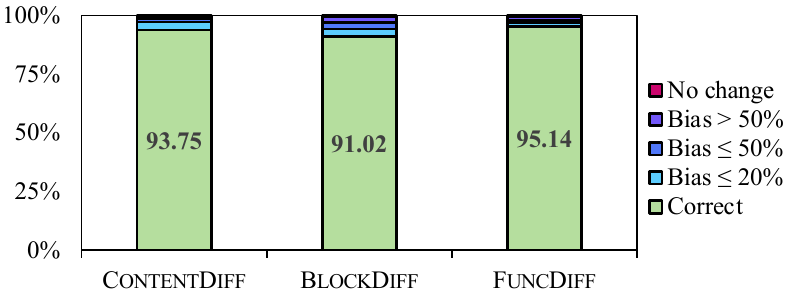}
    \caption{The accuracy of the format selection mechanism within \adaedit, trained on Qwen2.5-Coder-7B.}
    \label{fig:adaptive-exp}
\end{figure}

For interactive coding assistants, both latency and accuracy are critical user-centric metrics. 
As illustrated in Figure~\ref{fig:cost-effect}, we observe that purely diff-based formats do not provide a substantial latency advantage, especially \contentdiff whose average latency even exceeds that of \fullcode.
This counter-intuitive result is largely attributed to the characteristics of current benchmarks, which primarily consist of short, function-level samples.
In such contexts, the anchor content required by diff formats constitutes a high fraction of the total token count.
In contrast, \adaedit consistently shifts toward the top-left quadrant of the accuracy-latency spectrum, which not only reduces overall edit latency but also improves accuracy.

To dissect efficiency scaling properties, we analyze inference cost measured in the total output tokens. 
As depicted in Figure~\ref{fig:cost}, the cost analysis reveals three key insights: 
(\romannumeral1) The output tokens of \fullcode are directly proportional to the code length, making its cost prohibitive for long code. 
(\romannumeral2) The overhead from function-level anchors makes \funcdiff inefficient on short code, but its efficiency gradually surpasses \fullcode for longer code.
(\romannumeral3) \adaedit intelligently switches between both formats, which achieves optimal efficiency across all code scales and reduces cost by
over 30\% on long-code editing tasks. 

As efficiency gains are more significant for input code exceeding 300 tokens, we analyze accuracy and cost specifically for this subset of CanItEdit. 
The results in Table~\ref{tab:canitedit} indicate that despite a marked reduction in computational cost, edit accuracy does not suffer a significant decline for structure-aware formats. 
Notably, \funcdiff and its combination with \adaedit even exhibit marginal accuracy improvements, which suggests a promising solution to the tension between accuracy and efficiency in long-code editing scenarios.

\subsection{Analysis of the Adaptive Strategy}

\begin{figure}
    \centering
    \includegraphics[width=\linewidth]{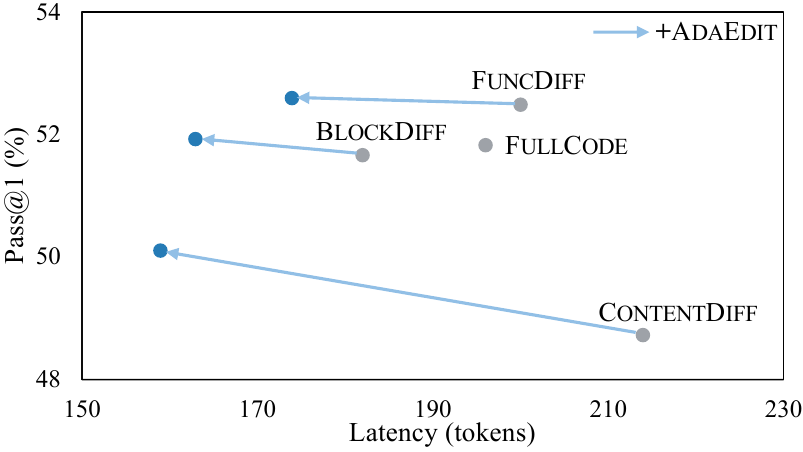}
    \caption{Latency-accuracy landscape, trained on InstructCoder with Qwen2.5-Coder-7B.}
    \label{fig:instructcoder-scatter}
\end{figure}

While \adaedit demonstrates strong performance in both effectiveness and efficiency, we further isolate and evaluate the reliability of its format selection mechanism using comprehensive categories. 
The ``No change'' category denotes instances where the model produces no valid edits. 
This typically occurs when the model directly echoes the input code without modifications, or generates invalid diffs. 
Furthermore, the ``Bias'' categories quantify the degree of token count deviation when the model makes an incorrect format selection.

Figure~\ref{fig:adaptive-exp} shows that the accuracy of selecting the most token-efficient format exceeds 90\% across all diff formats. 
When allowing for a slight deviation of 20\% in token counts, the average accuracy surpasses 95\%. 
Furthermore, we observe that selection correctness is closely related to the complexity of the diff representation. 
Simpler formats typically yield higher selection accuracy, whereas the decision-making burden is heavier for \blockdiff due to its inclusion of diverse control structures. 

To further validate the necessity of our adaptive strategy, we evaluate stronger LLMs \cite{deepseek3p2,gpt5} under a few-shot, inference-only setting. 
As detailed in Appendix~\ref{appendix:large_models}, these advanced models exhibit poor format selection accuracy and severe bias. 
These contrasting results confirm that LLMs do not inherently possess this cost-benefit logic, and that \adaedit is essential for effectively internalizing it.

\subsection{Other Training Corpus}

We conduct additional experiments using another Python training set, as shown in Figure~\ref{fig:instructcoder-scatter} and detailed in Appendix~\ref{appendix:corpus}. 
Due to inherent differences in data quality, the absolute accuracy scores are lower than those observed with OCEData. 
However, the comparative relationships regarding accuracy and efficiency remain highly consistent across all edit formats. 
This stability demonstrates that the advantages of our designs can generalize effectively across different training data.

\begin{table}
    \centering
    {\small
    \begin{tabular}{l|c}
    \toprule
    
    Formats & HumanEvalFix-JavaScript \\
    
    \midrule

    Base model              & 63.48 \\
    \fullcode         & \underline{66.13} \\

    \midrule
    
    \contentdiff        & 56.55 \\
    \quad w/ \adaedit   & 64.97 \\

    
    \blockdiff          & 62.44 \\
    \quad w/ \adaedit   & 65.70 \\
    
    \funcdiff           & 63.84 \\
    \quad w/ \adaedit   & \textbf{67.74} \\

    \bottomrule
    \end{tabular}}

    \caption{Pass@1 comparison, trained on the JavaScript subset of CommitPackFT with Qwen2.5-Coder-7B.}
    \label{tab:js-exp}
\end{table}

Our structure-aware formats can be readily extended to other programming languages such as JavaScript by simply adjusting the AST node configurations. 
The results in Table~\ref{tab:js-exp} demonstrate that \blockdiff and \funcdiff exhibit an even more pronounced advantage over the line-level \contentdiff in JavaScript. 
When integrated with \adaedit, the edit accuracy matches and even surpasses that of full-code generation. 
This highlights the cross-language robustness of our designs and the potential for multilingual code editing.

\section{Conclusion}

In this paper, we address the critical efficiency bottleneck in LLM-based code editing caused by the prevalent full-code generation paradigm. 
We begin with a systematic study that reveals the inherent unnaturalness of conventional diff formats for LLM generation.
To resolve this, we introduce \blockdiff and \funcdiff, two structure-aware diff formats that represent changes as block-level rewrites. 
Recognizing that any single format is not universally optimal, we then propose \adaedit, a general adaptive edit strategy that trains LLMs to dynamically choose the most token-efficient format for each editing task. 
Extensive experiments show that \adaedit paired with structure-aware diff formats reduces both latency and cost by over 30\% on long-code editing tasks, while not compromising accuracy.
Our work highlights the focus on the edit format itself and paves the way for more cost-effective coding assistants.

For future work, we identify two promising directions. 
First, while \adaedit establishes a robust cold-start foundation through supervised fine-tuning on the token-efficient pattern, incorporating reinforcement learning from verifiable rewards \cite{Li2022AlphaCode} could allow models to autonomously explore and optimize intrinsic editing formats by balancing functional correctness and token efficiency in the reward function. 
Second, we plan to explore fluid granularity that transcends fixed AST boundaries. 
Although implementing semantically aggregated changes requires significantly more complex diff and patch architectures, it represents a highly valuable step toward more flexible and intelligent code editing.

\section*{Acknowledgments}

This work was supported by the National Natural Science Foundation of China (No. 62272219) and the Fundamental and Interdisciplinary Disciplines Breakthrough Plan of the Ministry of Education of China (No. JYB2025XDXM118).

\section*{Ethical Considerations}

The datasets, benchmarks, and LLMs used in this work are public with permissive licenses.

\section*{Limitations}

Our work has several limitations that also frame promising directions for future research.

The effectiveness of our proposed formats is closely tied to the capabilities of the base model. 
Unlike full-code generation, interpreting edit intention to produce a structure-aware diff or adaptively choosing the optimal format is more sophisticated. 
Our results confirm this dependency, as the performance benefits of our formats consistently scale with model size and capability.

Furthermore, our study is constrained by the general scarcity of high-quality, large-scale training data for complex code editing. 
Developing datasets for long-code and even repository-level editing \cite{Bairi2024CodePlan,swebench} to scale edit formats remains a crucial undertaking.

As with all diff-based representations, including our structure-aware formats, there exist inherent risks of patch application failures or unintended code corruption. 
These potential issues necessitate careful validation when deploying such systems in real-world production environments.
We mitigate these concerns in Section~\ref{sec:effect} by verifying this basic edit usability for structure-aware formats.

\bibliography{custom}


\appendix

\section{Styles of Diff Formats}
\label{appendix:style}

\begin{figure}
    \centering
    \includegraphics[width=\linewidth]{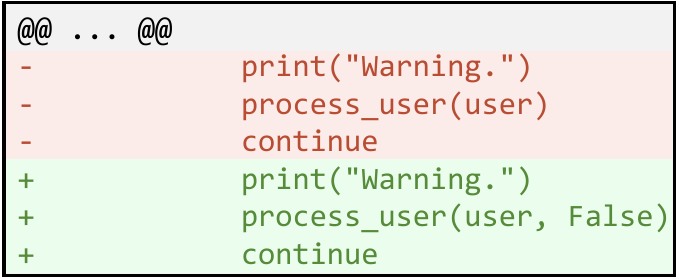}
    \caption{An example of \mincontentdiff using the hunk rewrite style.}
    \label{fig:example-rewrite}
\end{figure}

\begin{figure}
    \centering
    \includegraphics[width=\linewidth]{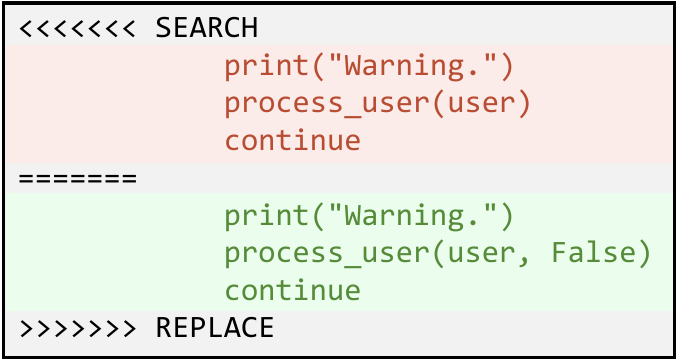}
    \caption{An example of \mincontentdiff using the search/replace style.}
    \label{fig:example-search}
\end{figure}

We provide an investigation into three hunk styles for content-addressed diff formats, including: 
\begin{itemize}
    \item the hunk rewrite style, which directly specifies the target code block to be rewritten. An example is shown in Figure~\ref{fig:example-rewrite}.

    \item the unified diff-like style, which incorporates context lines alongside addition and deletion markers.
    An example is shown in Figure~\ref{fig:example-diffs}\ref{fig:example-mincontentdiff}.

    \item the search/replace style \cite{aider2023edit}, which utilizes specific delimiters to separate the original block from its replacement.
    An example is shown in Figure~\ref{fig:example-search}.
\end{itemize}

Our experimental results in Table~\ref{tab:style-exp} indicate that the unified diff-like style suffers from a significant decline in edit accuracy. 
This performance drop occurs since the inclusion of unchanged context lines often disorients the model compared to the simpler binary logic of addition and deletion. 
Furthermore, while the search/replace style and the hunk rewrite style demonstrate comparable accuracy, the former introduces a critical reliability risk. 
Specifically, the search/replace style can fail to patch if the source code contains characters identical to its special delimiters. 
This potential for catastrophic failure constitutes a direct violation of the reliability requirements established in Equation~\ref{eq:format}. 
Consequently, we adopt the hunk rewrite style for all content-addressed diff formats to ensure maximum robustness and consistency across diverse scenarios.

\section{Details of Experiment Setup}

\subsection{Details of Datasets}
\label{appendix:datasets}

We include three training datasets:
\begin{itemize}
    \item \textbf{OCEData} \cite{ocedata}, a high-quality, synthetic Python dataset for code editing.
    Its samples cover a wider range of difficulty levels and features descriptive and lazy instruction styles.
    We use the full set for enough samples to adapt edit formats.
    
    \item \textbf{InstructCoder} \cite{instructcoder}, a synthetic Python dataset based on the Self-Instruct framework \cite{Wang2023selfinstruct}. 
    It consists mainly of short, function-level snippets for general-purpose code editing.

    \item \textbf{CommitPackFT} \cite{humanevalfix}, a real-world multilingual dataset filtered from GitHub commits.
    However, its generally limited quality and the lack of informative commit messages yields limited improvements in model training \cite{Aggarwal2025NextCoder}.
    We chose the JavaScript subset since it has the most samples among mainstream programming languages other than Python.
\end{itemize}

\begin{table*}
    \centering
    {\small
    \begin{tabular}{l|ccccc|c}
    \toprule
    
    Formats & EditEval & CanItEdit  & HumanEvalFix  & Aider-1  & Aider-2  & Average \\
    
    \midrule

    \mincontentdiff     & \textbf{61.42} & \textbf{37.02}  & \textbf{52.20}  & 35.56  & 42.22  & 45.68 \\
    \quad w/ interlaced    & 58.66 & 32.12  & 47.16  & 28.89  & 40.00  & 41.37 \\
    \quad w/ search/replace & 60.26 & 36.57  & 51.86  & \textbf{37.78}  & \textbf{42.96}  & \textbf{45.89} \\

    \midrule

    \contentdiff        & 67.91 & 47.19  & 65.91  & \textbf{43.70}  & 47.41  & 54.43 \\
    \quad w/ interlaced    & 56.11 & 31.57  & 44.63  & \textbf{43.70}  & 49.63  & 45.13 \\
    \quad w/ search/replace & \textbf{69.36} & \textbf{48.38}  & \textbf{66.95}  & 42.22  & \textbf{52.59}  & \textbf{55.90} \\

    \midrule
    
    \blockdiff        & 69.95 & \textbf{48.07}  & \textbf{64.85}  & \textbf{45.19}  & 51.85  & \textbf{55.98} \\
    \quad w/ interlaced    & 60.26 & 40.62  & 51.49  & 43.70  & 48.89  & 48.99 \\
    \quad w/ search/replace & \textbf{70.15} & 47.31  & 63.14  & 44.44  & \textbf{53.33}  & 55.68 \\
    
    \midrule
    
    \funcdiff         & \textbf{70.88} & \textbf{50.31}  & \textbf{67.62}  & \textbf{45.93}  & 51.85  & \textbf{57.32} \\
    \quad w/ interlaced    & 61.01 & 40.31  & 52.13  & 40.00  & 45.19  & 47.73 \\
    \quad w/ search/replace & 69.87 & 49.05 & 66.04 & 45.19 & \textbf{53.33} & 56.69 \\

    \bottomrule
    \end{tabular}}

    \caption{Style comparison of content-addressed diff formats on Qwen2.5-Coder-7B.}
    \label{tab:style-exp}
    
\end{table*}

To ensure data quality, we exclude samples whose output code contains syntax errors or whose edits yield no code changes. 
For Python datasets, we further format both the source and target code with Black \cite{black} to minimize the impact of non-semantic modifications.

We evaluate on four diverse Python code editing benchmarks. 
The first two instruction-guided benchmarks aligns with the training task:
\begin{itemize}
    \item \textbf{EditEval} \cite{instructcoder} is curated from GitHub commit data, MBPP \cite{mbpp}, and HumanEval \cite{humaneval}. 

    \item \textbf{CanItEdit} \cite{canitedit} is a hand-crafted benchmark featuring both descriptive and lazy editing instructions.
    Its samples frequently involve edits on longer code snippets.
\end{itemize}

The other two benchmarks involve bug-fixing tasks that do not provide typical instructions:
\begin{itemize}
    \item \textbf{HumanEvalFix} \cite{humanevalfix} is the Python subset of code repair scenario from HumanEvalPack, which requires models to fix a buggy function based on correct unit tests, without explicit edit instruction.
    
    \item \textbf{Aider} \cite{aider} simulates realistic Python coding exercises involving up to two attempts. 
    The first stage, \textbf{Aider-1}, is akin to code generation, where the model completes a starting Python file based on a rich problem description. 
    If the associated test suite fails, the second stage, \textbf{Aider-2}, asks the model to fix the previously generated buggy code based on the \texttt{pytest} \cite{pytest} results.
\end{itemize}

The statistics of training datasets and evaluation benchmarks are presented in Table~\ref{tab:dataset-stats}.
Moreover, we report the proportion of the samples where the diff format is selected by \adaedit in Table~\ref{tab:adaedit-qwen}.

\begin{table*}[t]
  \centering
  \resizebox{\textwidth}{!}{
  \begin{tabular}{l|ccc|ccccc}
    \toprule
    \multirow{2}{*}{Features} & 
    \multirow{2}{*}{OCEData} & 
    \multirow{2}{*}{InstructCoder} & 
    \multirow{2}{*}{\makecell{CommitPackFT-\\JavaScript}} & 
    \multicolumn{2}{c}{HumanEvalFix} &
    \multirow{2}{*}{EditEval} & 
    \multirow{2}{*}{CanItEdit} & 
    \multirow{2}{*}{Aider-1} \\
    \cmidrule(lr){5-6}
    & & & & Python & JavaScript & & & \\
    
    \midrule\
    \# Samples & 59,091 & 101,715 & 52,142 & \multicolumn{2}{c}{164} & 194 & 210 & 135 \\ 
    \midrule
    Avg. \# instruction & 104.86 & \ \ 20.20 & \ \ \ \ 8.59 & 220.52 & 220.63 & 20.50 & \ \ 52.71 & 534.73 \\ 
    Avg. \# input code & 206.73 & 108.07 & 193.63 & \ \ 63.85 & 100.98 & 72.02 & 317.76 & \ \ 56.51 \\ 
    Avg. \# reference & 301.23 & 151.01 & 214.38 & \ \ 64.51 & 102.30 & 87.75 & 381.70 & 242.67 \\ 
    Avg. \# diff lines & \ \ 23.62 & \ \ 13.10 & \ \ \ \ 9.13 & \ \ \ \ 2.21 & \ \ \ \ 2.40 & \ \ 8.68 & \ \ 13.82 & \ \ 34.76 \\ 
    \midrule
    Max. \# instruction & \ \ \,472 & \ \ \,112 & 134 & 1,217 & 1,309 & \ \ 73 & \ \ \,282 & 3,414 \\ 
    Max. \# input code & \ \ \,882 & \ \ \,729 & 663 & \ \ \,248 & \ \ \,432 & 235 & 1,544 & 2,133 \\ 
    Max. \# reference & 2,066 & 1,394 & 759 & \ \ \,247 & \ \ \,429 & 345 & 1,587 & 2,063 \\ 
    Max. \# diff lines & \ \ \,322 & \ \ \,211 & 127 & \ \ \ \ \ \ \,8 & \ \ \ \ \,10 & \ \ 46 & \ \ \ \ \,59 & \ \ \,364 \\ 
    \bottomrule
  \end{tabular}
  }
  \caption{Statistics of the training dataset and evaluation benchmarks, calculated by the Qwen2.5-Coder tokenizer.}
  \label{tab:dataset-stats}
\end{table*}

\begin{table}
    \centering
    \resizebox{\linewidth}{!}{
    \begin{tabular}{l|ccc}
    \toprule
    
    Formats & OCEData & InstructCoder & \makecell{CommitPackFT-\\JavaScript} \\
    
    \midrule

    \contentdiff    & 22.35 & 14.78 & 52.54 \\
    \blockdiff      & 24.07 & 21.79 & 31.58 \\
    \funcdiff       & 19.33 & 16.18 & 27.45 \\

    \bottomrule
    \end{tabular}}

    \caption{Proportion of the samples (\%) where the diff format is selected by \adaedit, using the Qwen2.5-Coder tokenizer.}
    \label{tab:adaedit-qwen}
\end{table}

\subsection{Details of Implementation Details}
\label{appendix:impl}

\paragraph{Training and Inference}
All models are finetuned for 3 epochs using the AdamW optimizer with a cosine scheduler \cite{Loshchilov2019Decoupled}, which has a peak learning rate of 3e-5 and a warmup ratio of 0.1.
Full-parameter SFT is conducted on 8 A100 (80GB) GPUs using a combination of Fully Sharded Data Parallel (FSDP2) \cite{Paszke2019PyTorch,Rajbhandari2020ZeRO} and sequence parallelism \cite{Li2023Sequence}, achieving an effective batch size of 256 with a maximum sequence length of 4,096 tokens.

For inference, we employ the vLLM library \cite{Kwon2023Efficient} and 4 A100 GPUs for efficient inference, with a maximum generation length of 4,096 tokens. 
We report the results of checkpoints with the best average accuracy.
Please refer to Appendix~\ref{appendix:metric} for detailed inference hyperparameters.

\paragraph{Prompt Template}
As shown in the text box, all our experiments employ a unified prompt template.
During Training, the prompt ends exactly at `\verb|### Response\n|'. 
The subsequent code or diff, which follows immediately, is used as the target sequence for calculating the cross-entropy loss.
During Inference, to guide the model's output format, we populate the prefix that acts as a starting sequence for generation:
\begin{itemize}
    \item For full-code generation, it is set to `\verb|```python\n|'.
    \item For any fixed diff format, it is set to `\verb|```diff\n|'.
    \item For \adaedit, we provide only `\verb|```|'. This minimal prefix intentionally leaves the format ambiguous, compelling the model to decide whether to use a diff format or full code by first producing the format specifier itself, followed by the content.
\end{itemize}

\begin{promptbox}{Prompt Template for Code Editing}
\begin{verbatim}
### Instruction
{INSTRUCTION}

### Input Code
```python
{INPUT_CODE}
```

### Response
```python/diff
{RESPONSE}
```
\end{verbatim}
\end{promptbox}

\begin{figure}
    \centering
    \includegraphics[width=\linewidth]{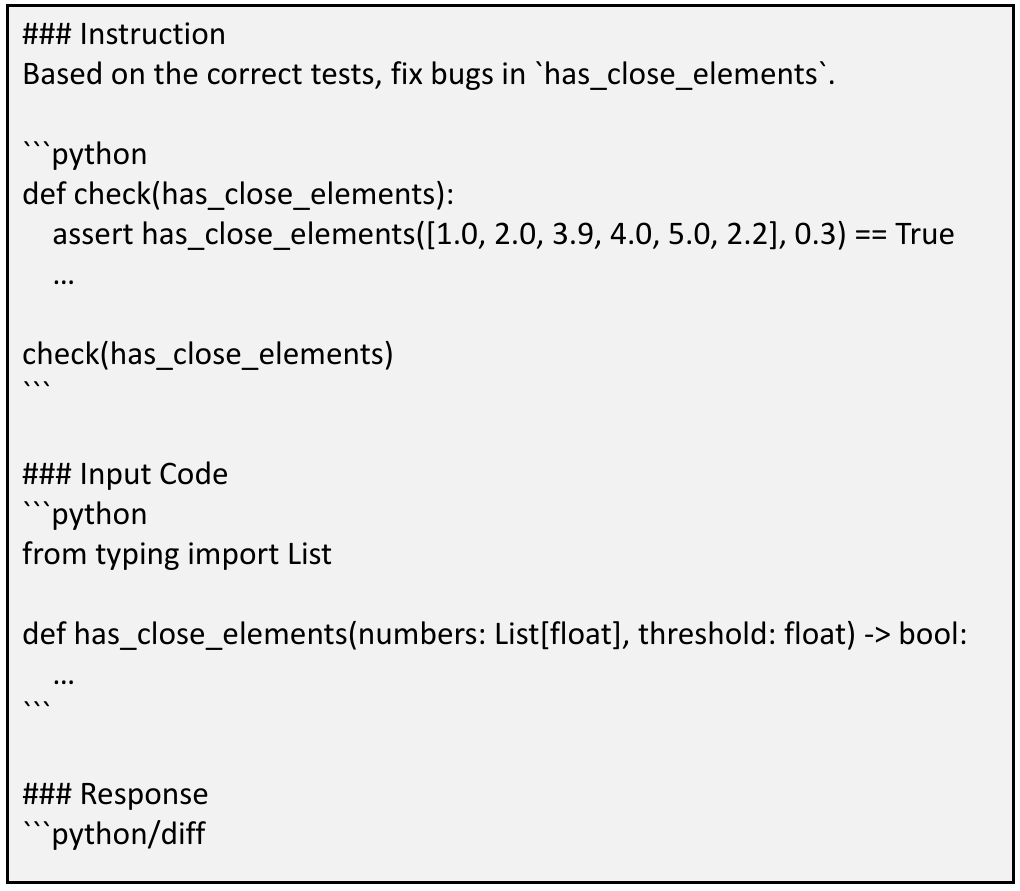}
    \caption{A prompt example of HumanEvalFix.}
    \label{fig:prompt-humanevalfix}
\end{figure}

\begin{figure}
    \centering
    \includegraphics[width=\linewidth]{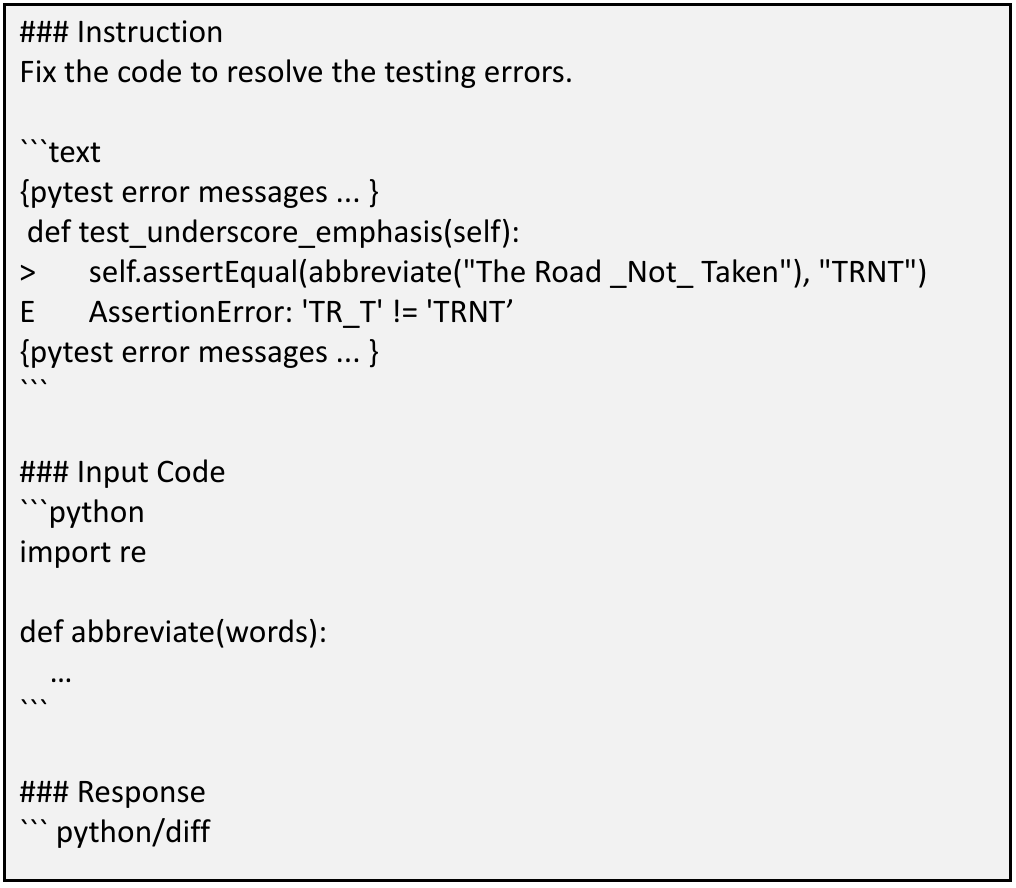}
    \caption{A prompt example of Aider-2.}
    \label{fig:prompt-aider2}
\end{figure}

For benchmarks \cite{instructcoder,canitedit} that provide explicit natural language instructions, we directly apply the above prompt template. 
For benchmarks \cite{humanevalfix,aider} that simulate bug-fixing scenarios without explicit instructions, we adopt a standardized approach to formulate the prompt:
\begin{itemize}
\item \textbf{HumanEvalFix:} the instruction is generated from a fixed template with the correct unit tests. 
An example is provided in Figure~\ref{fig:prompt-humanevalfix}.

\item \textbf{Aider-2:} we use a fixed instruction with the error messages from \texttt{pytest}. 
An example is shown in Figure~\ref{fig:prompt-aider2}.
\end{itemize}

\begin{table*}
    \centering
    {\small
    \begin{tabular}{c|l|ccccc|c}
    \toprule
    
    Base models & Formats & EditEval & CanItEdit & HumanEvalFix & Aider-1 & Aider-2 & Average \\
    
    \midrule

    \multirow{6}{*}{\makecell{DeepSeek-\\Coder-6.7B}} 
    & \fullcode             & 64.38 & 44.88  & 56.95  & 45.19  & 49.63  & 52.21 \\
    \cmidrule{2-8}
    
    & \minunidiff           & 11.13 & 5.38  & 5.88  & 5.93  & 8.89  & 7.44 \\
    & \quad w/ numbers      & 33.94 & 17.67  & 32.53  & 21.48  & 26.67  & 26.46 \\
    
    & \unidiff              & 37.89 & 9.10  & 26.80  & 28.15  & 30.37  & 26.46 \\
    & \quad w/ numbers      & 36.03 & 18.33  & 31.28  & 29.63  & 34.07  & 29.87 \\
    
    & \mincontentdiff       & 53.51 & 27.21 & 42.13 & 37.04 & 40.00 & 39.98 \\

    \midrule
    
    \multirow{6}{*}{\makecell{Qwen2.5-\\Coder-14B}} 
    & \fullcode             & 69.59 & 60.95  & 70.40  & 54.07  & 64.44  & 63.89 \\
    \cmidrule{2-8}
    
    & \minunidiff           & 29.10 & 11.62  & 19.15  & 24.44  & 31.11  & 23.08 \\
    & \quad w/ numbers      & 52.06 & 34.57  & 51.65  & 31.11  & 43.70  & 42.62 \\
    
    & \unidiff              & 53.22 & 21.50  & 50.58  & 37.78  & 46.67  & 41.95 \\
    & \quad w/ numbers      & 59.33 & 39.76  & 53.26  & 40.00  & 50.37  & 48.54 \\
    
    & \mincontentdiff       & 61.75 & 46.93  & 59.73  & 48.15  & 55.56  & 54.42 \\

    \bottomrule
    \end{tabular}}

    \caption{More results of conventional diff formats, trained on DeepSeek-Coder-6.7B and Qwen2.5-Coder-14B.}
    \label{tab:more-tra-exp} 
\end{table*}

\subsection{Details of Evaluation Metrics}
\label{appendix:metric}

The details of each metric are as follows:
\begin{itemize}
    \item \textbf{Accuracy:} we report the unbiased pass@1 score \cite{humaneval,Holtzman2020The}, using 20 samples with a temperature of 0.2 and a top-p value of 0.95.

    \item \textbf{Usability:} we report the exact patch-apply success rate, and the percentage of edited code without lint errors, using the popular Python linter \texttt{pylint} \cite{pylint}.

    \item \textbf{Latency:} we adapt the concept of First Meaningful Paint \cite{google2017FMP} from web performance to IDE latency.
    We measure it as the average number of tokens required to generate the first output that can be meaningfully rendered to the user.
    For full-code generation, the code is only renderable upon completion of the entire output. 
    For diff formats, this corresponds to the completion of the first diff hunk, which is the first moment a user can perceive a tangible change.

    \item \textbf{Cost:} we report the average number of tokens in the model's complete output, which corresponds to the overall computational cost of generating the edit.
\end{itemize}

For Aider, which involves multi-turn interactions, we report pass@1 using greedy sampling and omit efficiency comparisons for Aider-2 to ensure fairness. 

For \adaedit, we report the percentage of samples where the model correctly selects the most token-efficient edit format. 
To evaluate this selection accuracy, we convert each generated output of \adaedit into its alternative representation to facilitate a direct comparison. 
By calculating and comparing their token counts, we determine whether the model successfully identified the representation with the lower cost.

\section{More Results of Conventional Diff Formats}
\label{appendix:conventional}

As a supplement to Tables~\ref{tab:tra-exp} and~\ref{tab:main-exp}, we also evaluate conventional diff formats on DeepSeek-Coder-6.7B and Qwen2.5-Coder-14B.
The results in Table~\ref{tab:more-tra-exp} show that all number-indexed diff formats and \mincontentdiff yield edit accuracies far below that of the full-code baselines, which is consistent with conclusion on Qwen2.5-Coder-7B.

\begin{figure}
    \centering
    \includegraphics[width=\linewidth]{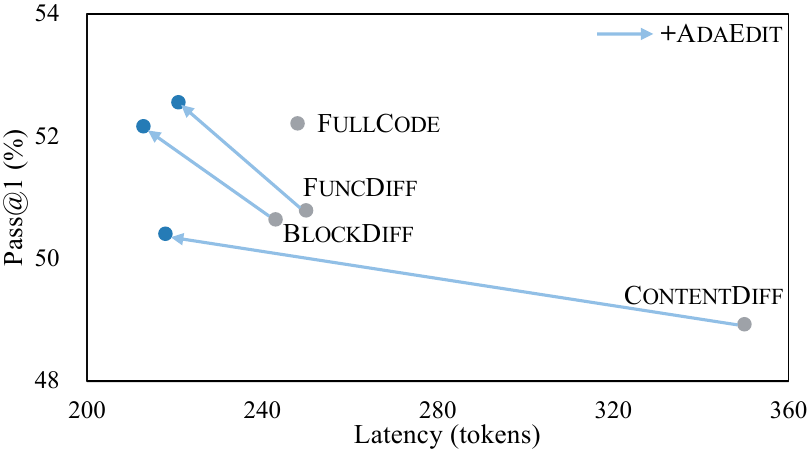}
    \caption{Latency-accuracy landscape of edit formats and \adaedit, trained on DeepSeek-Coder-6.7B.}
    \label{fig:ds-scatter}
\end{figure}

\begin{figure}
    \centering
    \includegraphics[width=\linewidth]{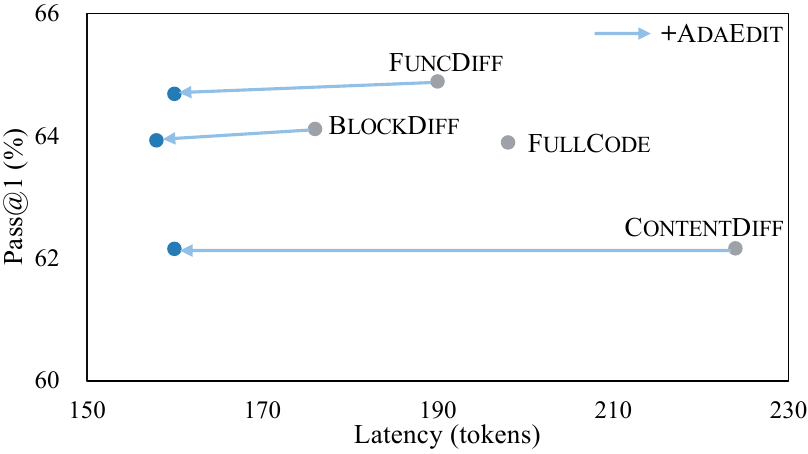}
    \caption{Latency-accuracy landscape of edit formats and \adaedit, trained on Qwen2.5-Coder-14B.}
    \label{fig:14b-scatter}
\end{figure}

\begin{figure}
    \centering
    \includegraphics[width=\linewidth]{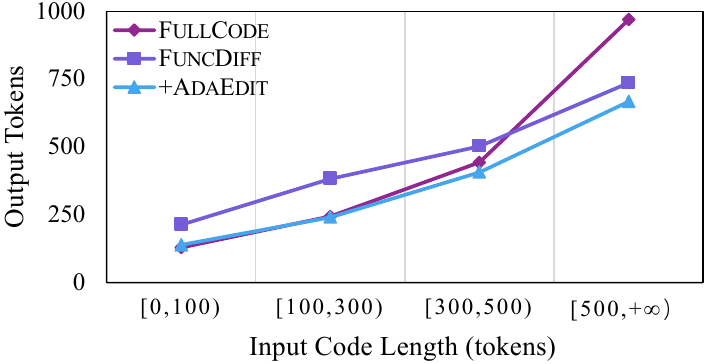}
    \caption{Edit cost comparison across code scales, trained on DeepSeek-Coder-6.7B.}
    \label{fig:ds-cost}
\end{figure}

\begin{table*}
    \centering
    {\small
    \begin{tabular}{l|ccccc|c}
    \toprule
    
    Formats & EditEval & CanItEdit  & HumanEvalFix  & Aider-1  & Aider-2  & Average \\
    
    \midrule

    Base model              & 55.39 & 42.98  & \textbf{65.12}  & 35.56  & 44.44  & 48.70 \\
    \fullcode         & 62.53 & \textbf{47.31}  & 61.89  & 39.26  & 48.15  & 51.83 \\

    \midrule

    \minunidiff         & 34.25 & 10.33  & 14.36  & 17.78  & 20.00  & 19.34 \\
    \quad w/ numbers    & 48.66 & 23.62  & 47.93  & 24.44  & 30.37  & 35.00 \\
    
    \unidiff            & 45.26 & 13.00  & 27.20  & 31.85  & 34.07  & 30.28 \\
    \quad w/ numbers    & 51.29 & 25.12  & 44.57  & 32.59  & 40.74  & 38.86 \\

    \midrule

    \mincontentdiff     & 54.12 & 32.60  & 53.57  & 31.11  & 35.56  & 41.39 \\
    
    \contentdiff        & 59.56 & 39.29  & 55.88  & \textbf{41.48}  & 47.41  & 48.72 \\
    \quad w/ \adaedit   & 62.16 & 35.05  & 62.90  & \underline{40.74}  & \underline{49.63}  & 50.10 \\

    \midrule
    
    \blockdiff        & 62.42 & 42.43  & 63.84  & 40.00  & \underline{49.63}  & 51.66 \\
    \quad w/ \adaedit   & \underline{63.79} & 43.57  & 61.89  & \underline{40.74}  & \underline{49.63}  & 51.92 \\
    
    \funcdiff         & \textbf{65.57} & 45.05  & \underline{64.42}  & 38.52  & 48.89  & \underline{52.49} \\
    \quad w/ \adaedit   & 63.04 & \underline{45.93} & 62.13 & \textbf{41.48} & \textbf{50.37} & \textbf{52.59} \\

    \bottomrule
    \end{tabular}}

    \caption{Pass@1 comparison on the InstructCoder dataset, trained on Qwen2.5-Coder-7B.}
    \label{tab:instructcoder-exp}
\end{table*}

\begin{figure}
    \centering
    \includegraphics[width=\linewidth]{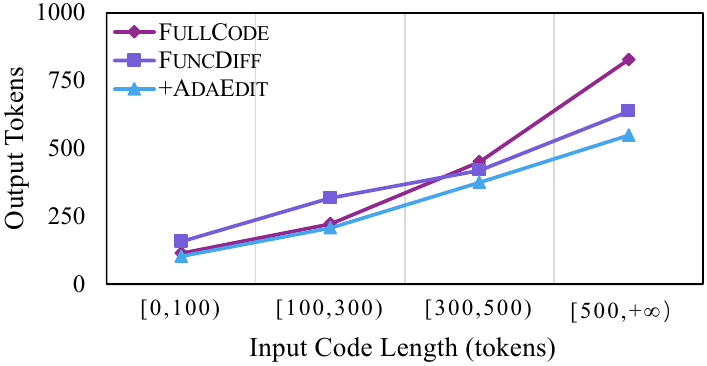}
    \caption{Edit cost comparison across code scales, trained on Qwen2.5-Coder-14B.}
    \label{fig:14b-cost}
\end{figure}

\section{More Results on Edit Efficiency}
\label{appendix:efficiency}

The additional evaluations of edit latency on DeepSeek-Coder-6.7B and Qwen2.5-Coder-14B are shown in Figures~\ref{fig:ds-scatter} and~\ref{fig:14b-scatter}, respectively.
Moreover, the additional evaluations of edit cost are shown in Figures~\ref{fig:ds-cost} and~\ref{fig:14b-cost}, respectively.
The results also demonstrate that \adaedit intelligently switches between both formats, which not only reduces overall edit latency but also achieves optimal efficiency across all code scales. 
These show the consistent conclusions with Qwen2.5-Coder-7B.

\section{Inference-only Evaluation on Larger Models}
\label{appendix:large_models}

\begin{table*}
    \centering
    {\small
    \begin{tabular}{l|ccccc}
    \toprule
    
    Models & Correct & Bias $\le$ 20\%  & Bias $\le$ 50\%  & Bias > 50\%  & No change \\
    
    \midrule

    DeepSeek-V3.2 & 25.34 & 4.97 & 4.81 & 7.03 & 57.85 \\
    
    Qwen2.5-72B-Instruct & 50.41 & 1.65 & 2.82 & 4.81 & 40.18 \\
    
    GPT-5 & 56.51 & 4.77 & 7.37 & 24.20 & 7.04 \\

    \midrule

    \adaedit & \textbf{91.02} & 3.07 & 2.73 & 2.50 & 0.67 \\

    \bottomrule
    \end{tabular}}

    \caption{The accuracy of the format selection mechanism using \blockdiff. \adaedit is finetued on Qwen2.5-Coder-7B, and other LLMs are evaluated in a few-shot,  inference-only setting.}
    \label{tab:inference-exp}
    
\end{table*}

While 7B-scale models are highly demanded for real-time coding tasks due to their inference efficiency, evaluating the capabilities of larger models is crucial from a research perspective to understand the broader applicability of \adaedit. 
To this end, we evaluate DeepSeek-V3.2 \cite{deepseek3p2}, Qwen2.5-72B-Instruct \cite{qwencoder}, and GPT-5-2025-08-07 \cite{gpt5} in a few-shot, inference-only setting. 
We design a system prompt instructing the models to choose the most token-efficient format between the full-code and \blockdiff formats based on the edit task.
It also includes a detailed description of the \blockdiff format alongside a single demonstration.

The format selection accuracy and bias distribution of these models are presented in Table~\ref{tab:inference-exp}. 
The results clearly indicate that all evaluated larger models perform suboptimally in selecting the most token-efficient formats. 
Even the highly capable GPT-5 achieves only a 56.51\% correctness rate, exhibiting a noticeable bias towards inappropriate format selection. 
Furthermore, DeepSeek-V3.2 and Qwen2.5-72B-Instruct exhibit extremely high ``No change'' rates. 
This phenomenon occurs since they struggle to learn the precise \blockdiff format from a single example and frequently default to generating standard unified diffs, which inevitably leads to application failures during the patching process. 
By comparison, \adaedit using Qwen2.5-Coder-7B achieves a remarkable 91.02\% accuracy rate. 
These findings strongly demonstrate that LLMs do not inherently possess the cost-benefit logic required for efficient format selection, further underscoring the necessity of \adaedit.

\begin{figure}
    \centering
    \includegraphics[width=\linewidth]{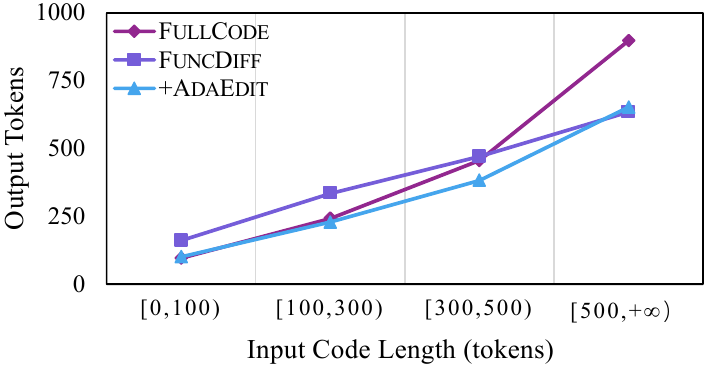}
    \caption{Edit cost comparison across code scales, trained on InstructCoder with Qwen2.5-Coder-7B.}
    \label{fig:instructcoder-cost}
\end{figure}

\section{More Results on InstructCoder}
\label{appendix:corpus}

To validate the generalizability of our structure-aware diff formats and \adaedit, we conduct additional experiments using the InstructCoder dataset with Qwen2.5-Coder-7B.

The results of edit accuracy are shown in Table~\ref{tab:instructcoder-exp}.
Evaluation results on latency and cost are illustrated in Figures~\ref{fig:instructcoder-scatter} and~\ref{fig:instructcoder-cost}, respectively.
Although variations in the training set influence the absolute values of the evaluation metrics, the relative performance and comparative trends among different edit formats remain consistent with the conclusions drawn from OCEData.

One outlier is the results on HumanEvalFix, where the base model outperforms all fine-tuned models.
We attribute the performance degradation to task shift on specific training data.
All edit formats are fine-tuned on edits driven by explicit natural language instructions (e.g., ``Replace function X with Y''). 
In contrast, samples in HumanEvalFix requires bug-fixing based on implicit information from unit tests (See Figure~\ref{fig:prompt-humanevalfix}). 
This shift favors the original capabilities of the base model, while the specialized, instruction-tuned models exhibit a slight performance degradation on this out-of-domain task. 
This is a common phenomenon in fine-tuning, called as catastrophic forgetting \cite{mccloskey1989catastrophic}.

Beyond specific task shifts, adapting LLMs to novel edit formats relies heavily on the scale and diversity of SFT data. 
However, the community faces a systemic scarcity of high-quality, verifiable training corpora. 
Existing benchmarks are predominantly constrained to Python function-level snippets, lacking coverage for other languages and complex repository-level scenarios \cite{Ma2025Lingma}. 
Therefore, constructing large-scale, diverse datasets and exploring accuracy-aware adaptive strategies stand as critical directions for future research \cite{Chen2024Neural,Luo2025Survey}.

\section{Time on Diff Generation and Patching}

The generation of our structure-aware diff formats is built upon \minunidiff, integrated with additional AST parsing and operations to ensure structural integrity. When combined with \adaedit, this process also includes the time required for text tokenization to facilitate format selection.

As shown in Table~\ref{tab:time_cost}, we record the time spent on diff generation and patching on the OCEData dataset, using \blockdiff combined with \adaedit.
Since diff generation is primarily conducted during the training data preparation phase, the computational overhead is highly efficient for large-scale data processing and can be readily parallelized across multiple CPUs. 
In contrast, the patching process during inference involves only straightforward text search and replacement operations. 
As evidenced by our timing analysis, the user-perceived patching latency is nearly instantaneous and does not negatively impact the overall user experience or system responsiveness.

Furthermore, we conduct a stress test on the diff generation process.
We process a massive Python file exceeding 10,000 lines, featuring complex class structures and scattered code modifications.
Calculating the \blockdiff for it takes merely 0.3 seconds, with the underlying block tree construction accounting for 0.28 seconds. 
These results demonstrate that the efficiency of structure-aware diff formats is highly robust, ensuring that computational overhead will not become a bottleneck for advanced inference techniques such as test-time scaling or future reinforcement learning pipelines.

\begin{table}
    \centering
    \resizebox{\linewidth}{!}{
    \begin{tabular}{l|ccc}
    \toprule
    
    Operations & Avg. time & Max. time & User impact \\
    
    \midrule

    Diff generation    & 3.40e-3 & 0.16 & - \\
    Patching           & 7.57e-5 & 7.19e-4 & Negligible \\

    \bottomrule
    \end{tabular}}

    \caption{Time spent on diff generation and patching in seconds, using \blockdiff combined with \adaedit on the OCEData dataset.}
    \label{tab:time_cost}
\end{table}

\end{document}